\documentclass[pre,twocolumn,showpacs,superscriptaddress,preprintnumbers,floatfix]{revtex4-1}

\usepackage{etex}
\usepackage{ifpdf}
\usepackage{hyperref}
\usepackage{dcolumn}
\usepackage{url}
\usepackage{amsmath}
\usepackage{amscd}
\usepackage{amsfonts}
\usepackage{amssymb}
\usepackage{bm}   
\usepackage{bbm}
\usepackage{verbatim}
\usepackage{stmaryrd}
\usepackage{amsthm}
\usepackage{xcolor}
\usepackage{setspace}
\usepackage{braket}

\usepackage{tikz}
\usetikzlibrary{arrows}
\usetikzlibrary{automata}
\usetikzlibrary{decorations.pathreplacing}
\usetikzlibrary{positioning}
\usetikzlibrary{plotmarks}
\usetikzlibrary{calc}
\usetikzlibrary{patterns}
\usetikzlibrary{external}
\usepackage{pgfplots}
\pgfplotsset{compat=newest}

\usepackage{graphicx}

\theoremstyle{plain}    
\theoremstyle{plain}    
\theoremstyle{plain}    
\theoremstyle{plain}    
\theoremstyle{plain}    
\theoremstyle{plain}    
\theoremstyle{plain}    
\theoremstyle{plain}    
\theoremstyle{plain}    
\theoremstyle{plain}    
\theoremstyle{plain}    
\theoremstyle{plain}

\newcommand{\CausalState}   { \mathcal{S} }

\newcommand{\hmu}       {h_\mu}

\newcommand{\forward}{+}
\newcommand{\reverse}{-}
\newcommand{\forwardreverse}{\pm} 

\newcommand{\FutureCausalState} { {\CausalState}^{\forward} }

\newcommand{\PastCausalState}   { {\CausalState}^{\reverse} }

\newcommand{\lastindex}[2]{
  \edef\tempa{0}
  \edef\tempb{#2}
  \ifx\tempa\tempb
    \edef\tempc{#1}
  \else
    \edef\tempa{0}
    \edef\tempb{#1}
    \ifx\tempa\tempb
      \edef\tempc{#2}
    \else
      \edef\tempc{#1+#2}
    \fi
  \fi
  \tempc
}

\newcommand{\CSjoint}[1][,]{
   \edef\tempa{:}
   \edef\tempb{#1}
   \ifx\tempa\tempb
      \ensuremath{\FutureCausalState\!#1\PastCausalState}
   \else
      \ensuremath{\FutureCausalState#1\PastCausalState}
   \fi
}

\newif\ifpm
\edef\tempa{\forwardreverse}
\edef\tempb{\pm}
\ifx\tempa\tempb
   \pmfalse
\else
   \pmtrue
\fi

\renewcommand{\H}{\operatorname{H}}

\begin{document}

\title{Memoryless Thermodynamics? A Reply}

\author{Dibyendu Mandal}
\email{dibyendu.mandal@berkeley.edu}
\affiliation{Department of Physics, University of California, Berkeley, CA
94720, U.S.A.}

\author{Alexander B. Boyd}
\email{abboyd@ucdavis.edu}
\affiliation{Complexity Sciences Center and Physics Department,
University of California at Davis, One Shields Avenue, Davis, CA 95616}

\author{James P. Crutchfield}
\email{chaos@ucdavis.edu}
\affiliation{Complexity Sciences Center and Physics Department,
University of California at Davis, One Shields Avenue, Davis, CA 95616}

\date{\today}
\bibliographystyle{unsrt}

\begin{abstract}
We reply to arXiv:1508.00203 `Comment on ``Identifying Functional
Thermodynamics in Autonomous Maxwellian Ratchets'' (arXiv:1507.01537v2)'.
\end{abstract}

\preprint{Santa Fe Institute Working Paper 15-08-XXX}
\preprint{arxiv.org:1508.XXXX [cond-mat.stat-mech]}

\maketitle



\section{Introduction}

Several years ago, Chris Jarzynski and one of us (DM) introduced a solvable
model of a thermodynamic ratchet that leveraged information to convert thermal
energy to work \cite{Mand012a,Lu14a}. Our hope was to give a new level of
understanding of the Second Law of Thermodynamics and one of its longest-lived
counterexamples---Maxwell's Demon. As it reads in ``bits'' from an input string
$Y$, a detailed-balance stochastic multistate controller raises or lowers a
mass against gravity, writing ``exhaust'' bits to an output string $Y'$.

A complete understanding of the ratchet's thermodynamics requires exactly
accounting for all of the information embedded the input and output strings and
how that information is changed by the ratchet. To simplify, we assumed the
input bits came from a biased coin and so the input information could be
measured using the \emph{single-bit Shannon entropy} $\H[Y_0]$. The information
in the output string was much more challenging to quantify, since correlations
are necessarily introduced by the action of the memoryful ratchet.
Unfortunately, due to mathematical complications arising from this, we could
only estimate the single-bit entropy $\H[Y'_0]$ of the output. Which, it must be said, is
only an upper bound on the actual information per output bit. Nonetheless, the
estimate of the change $\Delta \H = \H[Y'_0] - \H[Y_0]$ from input to output was
good enough to show that the ratchet was quite functional, operating as an
``engine'' in some regimes and an ``eraser'' in others.

Following in this spirit, the three of us here recently introduced a similar
memoryful ratchet for which all of the informational correlations in the output
bit string can be calculated exactly and in closed form \cite{Boyd15a}. As a
result, one of its contributions is that we could then show that the change
$d\hmu = \hmu[Y'] - \hmu[Y]$ in the \emph{Shannon entropy rate} $\hmu[X] =
\lim_{\ell \to \infty} \H[X_0 X_{1} \ldots X_{\ell}]/\ell$ allowed
one to identify all of the ratchet's thermodynamic functionality. We
emphasized, in particular, that using single-bit Shannon entropy
$\Delta \H$ would miss
much of that functionality, as $\H[Y'_0] \geq \hmu[Y']$. And, as such, we
generalized Refs. \cite{Bara2014b,Bara2014a} single-bit $\Delta \H$ ``Second
Law'' to use the Shannon entropy rate $d\hmu$. The underlying methods leveraged
a new way to account for the information storage and transformation induced by
memoryful channels \cite{Barn13a}. A similarly complete analytical treatment of
a companion Demon---Szilard's Engine---was recently given by two of us (AB and
JPC) \cite{Boyd14b}.

\section{Special Case of the Memoryless Transducer}

A recent arXiv post \cite{Merh15b} complained that our work~\cite{Boyd15a} is
misleading in certain aspects. It also claims priority over our entropy-rate
Second Law \cite[Eq. (4)]{Boyd15a}, stating that Eq. (24) of Ref.
~\cite{Merh15a} is the same. This is mathematically incorrect. Moreover, our
Ref. \cite{Boyd15a} is very clear about its contributions.  In
short, our treatment is more general, since it considers the much broader class
of Demons with arbitrary memory. Such Demons, as we describe in our manuscript,
can be represented as memoryful channels, otherwise known as transducers
~\cite{Barn13a}. In stark contrast, Ref. ~\cite{Merh15a}'s treatment is
sufficient only for describing memoryless channels; a highly restricted,
markedly simpler case. More to the point, its methods are inapplicable
to our memoryful channel setup. This error occurs in the proof of Ref.
\cite{Merh15a}'s Eq. (24) as it contains a statement that can be violated by
memoryful channels. We provide two counterexamples to this erroneous statement
in our response below. Finally, the case of memoryless Demons violates the
spirit of Refs. \cite{Mand012a,Lu14a}'s original work. The mathematical errors
and misinterpretation of physical relevance subvert the arXiv post's claims. We
now turn to respond to its three specific comments in greater detail.

\subsection*{Comment 1} 

In his first comment, the arXiv post's author mentions that the following
sentences in our paper give a {\it{``very strong misleading impression that the
paper above is the \textbf{first} to incorporate correlations successfully in
general"}} (using his own words). This is simple misreading, as our text makes
clear:
\begin{quote}
We introduce a family of Maxwellian Demons for which correlations among
information bearing degrees of freedom can be calculated exactly and in compact
analytical form. This allows one to precisely determine Demon functional
thermodynamic operating regimes, when previous methods either misclassify or
simply fail due to the approximations they invoke.
\end{quote}
Note that this explicitly mentions the solvable aspect of our model---that
correlations can be calculated exactly in a compact, analytical form. We stand
by the claim that ours is the first such solvable model. We did not claim to be
the first to consider correlations. More pointedly, the author's actual
article~\cite{Merh15a} \emph{does not have any model with calculable
correlations}. We justify the second sentence quoted above on identifying
functional thermodynamics through explicit calculations and diagrams in Sec. V
of our paper. The arXiv post ignores these.

The author claims that the {\textit{``main result in Section 4 of [1] was
exactly the same as the above mentioned upper bound on the extracted work in
terms of the change in the joint entropy ... ."}} In this, he refers to Eq. (4)
of our paper and claims that he had derived it before as Eq. (24) of his
paper~\cite{Merh15a}. While we agree that our equation superficially looks like
the infinite-time limit of the author's equation, their relationship is
different than a glance suggests:
\begin{itemize}
\item The author's proof of Eq. (24) \cite{Merh15a} does not apply to our setup.
This is because the author considered the much simpler case of memoryless
channels, whereas we considered the much more mathematically challenging case of memoryful channels. (We return to this point again in context of the $3^\text{rd}$ comment.)
\item Appendix A in our paper clearly shows that Eq. (4) there is valid only in
the asymptotic limit of stationary input bits for a finite-state Demon. (These
are standard assumptions in the field.) In absence of these assumptions, we
have a more general form of the Second Law discussed in detail in Appendix
A~\cite{Boyd15a}. The arXiv post neglects these discussions. 
\end{itemize}

\subsection*{Comment 2}

The arXiv post quotes the following from our paper:
\begin{quote}
In effect, they account for Demon information-processing by replacing the
Shannon information of the components as a whole by the sum of the components'
individual Shannon informations. Since the latter is larger than the former
[19], these analyses lead to weak bounds on the Demon performance.
\end{quote}
And, then goes on to claim that the second assertion may not be true if the
incoming bits $\{Y_i\}$ are correlated. This \emph{is} the case in
the author's
Ref.~\cite{Merh15a}, where the sum of individual entropy differences is
actually stronger, under the additional assumption that the Demon is
memoryless. We agree. But, as the author himself points out, our claim is true
if the incoming bits are uncorrelated. We explicitly state that we are
considering this case, where the input is uncorrelated, in the paragraph
following Eq. (4). And, this happens to be the case for all
the exactly solvable models of Maxwell's Demon developed so far (referred to by
``they" in the above quote). (We reiterate, the author has not given any
exactly solvable model of Maxwell's Demon with calculable correlations in~\cite{Merh15a}.)

The author did not sufficiently consider the remainder
of our development before expressing his criticism in public. After Eq. (4), we
explicitly mention the sufficient condition of uncorrelated incoming bits for
Eq. (4) to be stronger than Eq. (2).
 
According to the author \textit{``the point in second law and its
extensions ... should be to provide, first and foremost, an extended version of
the second law in a faithful manner, namely, to show the increase of the real
entropy of the entire system, including that of the information reservoir. In
the correlated case, the latter is given by the change in the joint entropy of
the symbols, regardless of whether or not this is smaller or larger than the
sum of individual entropy differences."} We disagree. This is nothing more
than an attempt to rewrite the history of physics.

The primary emphasis of the Second Law from its very inception has been on
the strongest possible bounds. When Sadi Carnot formulated the Second Law, it
was \textbf{all} about maximum efficiency of heat engines---the maximum
possible work that can be extracted \cite{Carn97a}. Entropy was a derived concept,
entering through the
works of Clausius and Thompson \footnote{Furthermore, entropy rate in the
information-theoretic sense is not always applicable in the thermodynamic
sense. This is seen in context of nonlinear, chaotic dynamics where information
(about the initial state) is continuously produced without any need for
thermodynamic irreversibility \cite{Dorf99a}. For thermodynamics, one must
consider the time-reversed description.}.

The author mentions that \textit{``bounds are useful when they are easier to
calculate than the real quantity of interest, which is not quite the case in
this context. Quite the contrary, joint entropies (especially of long blocks)
are much harder to calculate."} He fails to notice that we
attained precisely this ``hard" task by calculating exactly the entropy rate
$\hmu[Y'] = \lim_{\ell \rightarrow \infty} \H[Y'_{0:\ell}]/\ell$.
(We might, at this point, recommend the review of correlations and
information in random-variable blocks presented by Ref. \cite{Crut01a}.) And, the entropy rate is smaller than the individual entropy difference, which in our case has \textbf{observable consequences}, as discussed in detail in Sec. V of our paper. Even the later part of his comment \textit{``work itself ... depends only on the input and output marginals"} is not true in a generic memoryful situation. We have explicit examples (unpublished) where the extracted work also depends on correlations.

\subsection*{Comment 3}  

Here, the author claims that the \textit{``bound in [57] ... is exactly the
same as in eq. (4) of 1507.01537v2, except that in [57], no limit on is taken
over the normalized entropies (but this is because even stationarity is not
assumed there, so the limit might not exist). Moreover, while it is true that
in the model of [57] the channel was memoryless, the derivation itself of this
very same bound (in Section 4 of [57]) was not sensitive to the channel
memorylessness assumption."} (Citation [57] corresponds to Ref. \cite{Merh15a}
here.) We agree that Eq. (4) in our paper appeared in a somewhat different form
than in his paper, as Eq. (24). This is moot, however. His derivation
does not apply to our case nor to the original solvable Maxwell's demon ~\cite{Mand012a}. In his justification, the author says that the
\textit{``crucial step in [57] ... was the equality}
\begin{align}
\H(Y'_i | Y_1 , \ldots, Y_{i-1}, Y'_1, \ldots, Y'_{i-1})
  = \H(Y'_i | Y_1 , \ldots, Y_{i-1})
  ~,
\label{eq:FirstStep}
\end{align}
\textit{which is the case when} 
\begin{align}
Y'_i \rightarrow (Y_1 , \ldots, Y_{i-1}) \rightarrow (Y'_1 , \ldots, Y'_{i-1})
\label{eq:MarkovChain}
\end{align} 
\textit{forms a Markov chain, and this happens not only for a memoryless
channel, but for any causal channel without feedback, namely, 
\begin{align}
P(Y'_1 , \ldots, Y'_n | Y_1 , \ldots, Y_n )
  = \Pi_{i = 1}^n P (Y'_i | Y_1 , . . . , Y_i)
  ~.
\label{eq:CondlProb}
\end{align}
In physical terms, this actually means full generality."} 

This analysis is incorrect. Equation (\ref{eq:FirstStep}) above is not
sufficiently general, since it does not consider the case in which the Demon is
a memoryful channel. When the Demon has memory, its internal state can depend
on both the input past $Y_1 , \ldots, Y_{i-1}$ and output past $Y'_1 , \ldots,
Y'_{i-1}$. The Demon's internal states store information about the past of $Y$
or $Y'$ and can communicate it to the outgoing bits of $Y'$. The author's
assertion of ``full generality'' is false. In fact, the memoryless
assumption is violated for the original solvable model of Maxwell's
Demon ~\cite{Mand012a} in which the Demon has three internal states. For a
memoryful Demon Eq. (\ref{eq:MarkovChain}) above is not a Markov
chain. See Ref.~\cite{Barn13a}'s discussion of memoryful transduction.

To see how Eq. (\ref{eq:FirstStep}) can be violated, consider the case of a
memoryful Demon that simply ignores the input bits and outputs a period-$2$
process. This means that there are two possible output words:
\begin{align*}
\Pr(Y'_1Y'_2...=010101...) = \Pr(Y'_1Y'_2...=101010...) = 1/2
  ~.
\end{align*}
In this case the uncertainty of the $i$th output given the history of inputs is
$\H[Y'_i | Y_{1:i}]=1$, since we are completely uncertain as to
whether or not the $i$th bit is a zero or one.  Note that we used the notational shorthand $Y_{1:i}$ to represent the random variables $Y_1,Y_2, \ldots Y_{i-1}$. When we also condition
on the history of output bits, we find that we are completely certain of the
next bit, since we know the output's phase, and $\H[Y'_i | Y_{1:i}, Y'_{1:i}]=0$.  The most general relation for the
uncertainty of the output is:
\begin{align*}
\H[Y'_i | Y_{1:i}, Y'_{1:i}]
  \leq \H[Y'_i | Y_{1:i}]
  ~.
\end{align*}
This inequality, replacing Eq. (\ref{eq:FirstStep}) above, renders the proof in 
the author's paper inapplicable to our situation.

This reflects the fact that a memoryful ratchet can and typically does create
correlations among the outgoing bits even though the incoming bits may not be
correlated. In fact, we can exactly calculate the uncertainty in the next
output bit conditioned on the infinite length input and output histories of the
memoryful ratchet we describe in our Ref. \cite{Boyd15a}.  When the ratchet is
driven by a fair coin input process, the two quantities of interest are:
\begin{align*}
\lim_{i \rightarrow \infty}\H [Y'_i | Y_{1:i}]
  = \frac{1}{2}\left( \H\left(\frac{p}{2}\right)
  +\H\left(\frac{q}{2}\right)\right)
  ~,
\end{align*}
where $\H(b)$ is the binary entropy function for a coin of bias $b$
\cite{Cove06a},
and:
 \begin{align*}
 \lim_{i \rightarrow \infty} \H[Y'_i | Y_{1:i}, Y'_{1:i}]
   =\frac{1}{4}\left( \H\left(p\right)
 + \H\left(q\right)\right),
 \end{align*}
and their difference:
 \begin{align*}
 \lim_{i \rightarrow \infty} & (\H[Y'_i | Y_{1:i}]
   - \H[Y'_i | Y_{1:i}, Y'_{1:i}]) \\
 & = \frac{1}{2}\left( \H \left( \frac{p}{2}\right) -\frac{\H(p)}{2}\right)
   + \frac{1}{2}\left( \H \left( \frac{q}{2}\right) -\frac{\H(q)}{2}\right) \\
 & \geq 0
  ~,
\end{align*}
by the concavity of $\H(\cdot)$. This is only zero when p=q=0.
Thus, the assumption made in Eq. (1) is not just insufficiently general, but it is explicitly violated in the physical memoryful ratchet considered in our work.

On a more conceptual level, Eq. (4) in our paper is valid only in the
asymptotic limit of a stationary input with a finite-state Demon. Otherwise,
there would be natural generalizations incorporating the Demon's entropy and
its correlations with the bits, as is amply discussed in Appendix A of our
paper.

\section{Summing Up}

As our response to the arXiv post's Comment 3 just made plain, the essential
issue reduces to the post's author misapplying results for memoryless channels.
Most directly, the post's claim to priority for our entropy-rate Second Law is
invalid. Perhaps the simple memoryless channel case, one very broadly adopted
in elementary information theory \cite{Cove06a}, prevented the post's author
from appreciating this and related technical points. Whatever the motivation,
it led to the post's public airing of a series of grievances---grievances that
derive not from misleading text, but from the author's misinterpretation. That
said, we do appreciate the opportunity to emphasize the central role of memory
and structure in thermodynamics.

\section*{Acknowledgments}

As an External Faculty member, JPC thanks the Santa Fe Institute for its
hospitality during visits. This work was supported in part by the U. S. Army
Research Laboratory and the U.S. Army Research Office under contracts
W911NF-13-1-0390 and W911NF-12-1-0234.

\bibliography{chaos}

\end{document}